\documentclass[twocolumn,
showpacs,floatfix,prl
]{revtex4}
\usepackage{graphicx,amsmath,amssymb,amsbsy,latexsym,amsfonts,subfig
ure,bbm}
\usepackage[usenames]{color}
\usepackage{epic}
\usepackage{epsfig}
\newcommand{\AB}{$AB$-}
\newcommand{\vecrep}{\mathbf}

\newcommand{\Eins}{\mathbbm{1}}
\newcommand{\x}{\mathbf{x}}

\newcommand{\ket}[1]{\left| #1 \right>}
\newcommand{\bra}[1]{\left< #1 \right|}

\begin{document}
\title{Transport through graphene nanoribbons:\\ suppression of
  transverse quantization by symmetry breaking} \author{Florian
  Libisch} \email{florian.libisch@tuwien.ac.at}
\affiliation{Institute for Theoretical Physics, Vienna University of
  Technology\\Wiedner Hauptstra\ss e 8-10/136, A-1040 Vienna, Austria,
  European Union} \author{Stefan Rotter} \affiliation{Institute for
  Theoretical Physics, Vienna University of Technology\\Wiedner
  Hauptstra\ss e 8-10/136, A-1040 Vienna, Austria, European Union}
\author{Joachim Burgd\"orfer} \affiliation{Institute for Theoretical
  Physics, Vienna University of Technology\\Wiedner Hauptstra\ss e
  8-10/136, A-1040 Vienna, Austria, European Union} 

\begin{abstract}
We investigate transport through nanoribbons in the presence of
disorder scattering. We show that size quantization patterns are only
present when $SU(2)$ pseudospin symmetry is preserved. Symmetry
breaking disorder renders transverse quantization invisible, which may
provide an explanation for the necessity of suspending graphene
nanoconstrictions to obtain size quantization signatures in very
recent experiments. Employing a quasi-classical Monte-Carlo
simulation, we are able to reproduce and explain key qualitative
features of the full quantum-mechanical calculations.
\end{abstract}

\pacs{73.23.-b, 73.63.-b, 73.40.-c}

\maketitle


Graphene, the first two-dimensional solid \cite{MasslessFirsov},
features remarkable electrical and mechanical properties that open the
possibility for many new and intriguing applications
\cite{guinea_review} including high precision mechanical or chemical
sensors, ultrafast single-electron transistors, and spintronic
devices. Graphene quantum dots can meanwhile be fabricated with
well-defined dimensions \cite{pon08,gue08} allowing for the observation
of Coulomb blockade \cite{sta08ab,Guettinger09}, and Klein
tunneling-related phenomena \cite{kat06}.

\begin{figure}[]
\hbox{}\hfill\epsfig{file=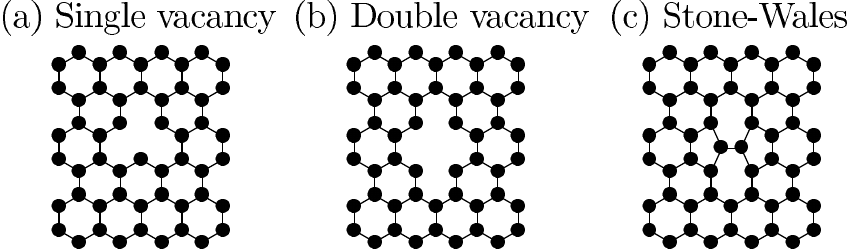,width=8cm}
\hfill\hbox{}
\caption{ Defects in graphene nanoribbons (a) single vacancy breaking
  the \AB sublattice symmetry, (b) double vacancy preserving the
  sublattice symmetry, (c) Stone-Wales deformation: four hexagons are
  replaced by two pentagons and two heptagons which also breaks \AB  
  sublattice symmetry.  
}
\label{fig:latt_defe}
\end{figure}

While properties of the perfect honeycomb lattice and its consequences
for electronic structure and dynamics are well understood, the effect
of disorder caused by local distortions of the lattice symmetry,
graphene-substrate interaction, or charged impurities on the unique
properties of graphene remains the focus of theoretical investigations
\cite{WimmerEdgestates, LewenkopfConduct, Ihnatsenka,
  LibischLandau}. On general grounds, local symmetry breaking in
low-dimensional systems is expected to have more pronounced effects
than in 3D bulk materials. In this Letter, we demonstrate the
remarkably strong sensitivity of quantum transport through graphene
nanoribbons to very low concentrations of point defects that break the
SU(2) pseudospin symmetry associated with the triangular
sublattices of graphene.
We simulate transport through
disordered nanoribbons of realistic size described by a tight binding
Hamiltonian. To interpret our numerical results we use
two simpler models: a continuous Dirac-like equation and a
quasi-classical Monte-Carlo simulation. We identify
pseudospin non-conserving scattering at lattice vacancies to be the
key for the breakdown of size quantization.


The ideal, infinitely extended graphene sheet features a honeycomb
lattice made up of two interleaved triangular sublattices (A and B).
It can be described in tight-binding approximation by
the Hamiltonian \cite{Wallace}
\begin{equation}
  H = \sum_{i,s}\ket{\phi_{i,s}}V_i\bra{\phi_{i,s}}-
  \sum_{(i,j),s}\gamma_{i,j}\ket{\phi_{i,s}}\bra{\phi_{j,s}} +
  h.c.\,, \label{H_Graph_TB}
\end{equation}
where the sum $(i,j)$ extends over pairs of lattice sites,
$\ket{\phi_{j,s}}$ is the tight-binding orbital with spin $s$ at
lattice site $j$, $V_i$ is a locally varying potential, and
$\gamma_{i,j}$ is the hopping matrix element between lattice sites $i$
and $j$. For improved accuracy, we describe the hexagonal graphene
lattice using third-nearest-neighbor coupling (for details see
Ref.~\onlinecite{libisch_edge}). We are interested in the effects of
bulk scattering at localized point defects and consider samples of
ribbons with an average width of up to 60 nm corresponding to $\approx
300$ unit cells in transverse direction ($y$) orthogonal to the
direction of transport ($x$). We use an approach suitable for the
efficient description of large-scale graphene nanodevices employing a
variant of the modular recursive green's function method (MRGM)
\cite{libisch_edge, Rott06}. Different defects (see
Fig.~\ref{fig:latt_defe}) can be easily included at module
boundaries. We perform ensemble averages over, typically, 100
different disorder realizations to eliminate any non-generic features.

We analyze the implications of the numerical results with the help of
simpler models. One of them employs the Dirac-like bandstructure of
graphene.  Close to the Fermi energy, the band structure of
Eq.~(\ref{H_Graph_TB}) can be approximated (assuming that $V_i \ll
\gamma_{i,j}$) by a conical dispersion relation around the $K$
point \cite{Semenoff},
\begin{equation}\label{Eq:Semenoff}
  E(k + k_K) = E(k_K) + k \partial_k E(k_K) +
  \mathcal{O}(k_K^2)\approx v_{\mathrm{F}}|k|,
\end{equation}
with the Dirac-like Hamiltonian,
\begin{equation}\label{Eq:H_Dirac}
H = \hbar v_{\mathrm F}\left(\begin{array}{cc}
0 & \partial_x + i \partial_y\\\partial_x - i \partial_y & 0
\end{array}\right) + \Eins\cdot V(\x),
\end{equation}
where we have set $E(k_K) = 0$. Equation~(\ref{Eq:H_Dirac}) ignores
both the length scale of the graphene lattice constant
$a=1.4$ \AA\ and the broken rotational symmetry of the cone due to the
hexagonal lattice structure, an effect known as triangular
warping \cite{McCann06b, guinea_review}. Eigenfunctions of
Eq.~({\ref{Eq:H_Dirac}) on an infinitely extended sheet $(V=0)$
  are plane waves $\ket k$ where the direction of motion
  $\theta_k$,
\begin{equation}\label{Eq:Thetak}
\theta_k = \tan^{-1}(k_y / k_x),
\end{equation} 
 is coupled to the \AB sublattice degree of freedom \cite{guinea_review},
\begin{equation}\label{pseudospin}
\ket{\vecrep{k}}= e^{i \vecrep{k\cdot r}}\left(\ket{A} + e^{i \theta_k}\ket{B}\right)/\sqrt 2.
\end{equation} 
The Hamiltonian $H$ preserves the SU(2) pseudospin projection (or
helicity) 
$h = (\vecrep{\boldsymbol{\sigma}}\;\cdot\;\vecrep k)/\left|k\right|$
 along $\hat{k}$, i.e.~the angle $\theta_k$ ($\boldsymbol\sigma$ is the
vector of the Pauli matrices). Furthermore, the band
 structure features \emph{two} non-equivalent cones (``valleys'') at
 the $K$ and $K'$ points in the reciprocal lattice. This additional
 degeneracy allows to represent the low-energy band-structure
 near $E=0$ in terms of Dirac-like four-spinors
 $\ket\psi=(\psi_A^{K},\psi_B^{K}, \psi_A^{K'},\psi_B^{K'})$
 with amplitudes for the \AB sublattice in real space and for the $KK'$ 
 points in reciprocal space. The sign of $\theta_k$ 
is reversed upon transition from $K$ to $K'$. Note that physical spin
 is not included in the present analysis.

One of the consequences of the preservation of pseudospin is the
suppression of backscattering \cite{guinea_review}. If the scattering
potential commutes with the helicity operator, the first-order
transition probability $P$ for scattering $\ket{\vecrep k}\rightarrow
\ket{\vecrep k'}$ is proportional to
\begin{equation}
P(\vecrep{k}\rightarrow\vecrep{k}') =\left|\bra{\vecrep
  k'}V\ket{\vecrep k}\right|^2 \propto \cos^2[(\theta_{\vecrep k} -
  \theta'_{\vecrep k'})/2],\label{proptozero}
\end{equation}
which vanishes for $\left|\theta_{\vecrep k} -
\theta'_{\vecrep k'}\right| = \pi$, i.e.~for backscattering. For
locally broken \AB sublattice symmetry [Eq.~(\ref{pseudospin})], the
pseudospin is no longer conserved and backscattering becomes
possible. Consequently, differential cross sections for scattering at
local defects that preserve the SU(2) pseudospin symmetry should obey
Eq.~(\ref{proptozero}) while defects that locally destroy the
\AB sublattice allow for scattering in arbitrary direction, in
particular for isotropic $s$-wave scattering in the long-wavelength
limit ($k\rightarrow 0$),
\begin{equation}
P(\vecrep{k}\rightarrow\vecrep{k}') \propto \mathrm{const}.\label{swavescatt}
\end{equation}
To elucidate the consequences of these specific scattering features we
incorporate them in a quasi-classical transport simulation based on
the propagation of Monte-Carlo ensembles of classical trajectories:
pseudospin conserving lattice defects are simulated by elastic
scattering probabilities of the form of Eq.~(\ref{proptozero}),
pseudospin non-conserving defects will be represented by isotropic
$s$-wave scattering [Eq.~(\ref{swavescatt})].  We randomly shoot
trajectories that move classically (i.e.~on straight lines) in between
scattering events. After traversing a mean free path $\lambda_s$
(determined by the disorder concentration), a scattering event with either
pseudospin conserving or non-conserving angular differential
scattering probability takes place.  Note that the only quantum input
are the differential scattering probabilities Eq.~(\ref{proptozero})
and Eq.~(\ref{swavescatt}).  A trajectory is counted as transmitted
(reflected) if it traverses the length $L \gg \lambda$ (returns past
the starting point).

As initial condition we choose the longitudinal wave numbers $k_{x,n}$  
corresponding to the quantized open modes $n$, for each energy $E$
\begin{equation}
 k_{x,n} = \sqrt{\left(\frac{E}{\hbar v_F}\right)^2  - 
\left(\frac{n\pi}{W}\right)^2},\quad n \in \mathbb{Z}.\label{graph_scatt}
\end{equation}
We use, typically, 200.000 trajectories per open channel with initial
momenta chosen according to Eq.~(\ref{graph_scatt}).  If the mismatch
$\left|k_y - k_{y,n}\right|$ between the transverse wave number $k_y$
and $k_{y,n}$ corresponding to the largest flux-carrying mode
(i.e.~the largest $n$ for which $k_{x,n}$ is real) is larger than
$\left|k_y - k_{y,n+1}\right|$ , scattering into an evanescent mode
(i.e.~complex $k_x$) is assumed, initiating a new scattering event in
backwards direction $(-k_x,k_y)$. The average over ensembles of
trajectories provides a quasi-classical Monte-Carlo estimate for the
conductance as a function of energy.


We consider three different lattice defects (Fig.~\ref{fig:latt_defe})
which locally perturb the electronic structure and, thus, introduce
disorder. The simplest defect is a point defect residing on a single
carbon atom. Such a defect can be caused, for example, by chemical
absorption of hydrogen, forming a covalent bond with the $p_z$ orbital
of a carbon atom, locally changing the electronic configuration from
$sp^2$ to $sp^3$. Consequently, the $p_z$ orbital of this carbon atom
no longer contributes to the electronic bandstructure of graphene. In
a tight-binding approximation, this can be modeled by a single
(electronic) lattice vacancy, i.e.~one carbon atom is effectively
removed from the graphene lattice [see
  Fig.~\ref{fig:latt_defe}(a)]. We consider an ensemble average
over many configurations of randomly placed point defects with a
relative defect density as small as $n_i=10^{-5}$ impurities/carbon.
As a second class of defects we consider double vacancies, i.e.~we
remove both atoms of a unit cell [see Fig.~\ref{fig:latt_defe}(b)]. In
contrast to a single vacancy, this defect does not break the \AB
sublattice symmetry. Double vacancies are pseudospin conserving as
both the A and the B lattice are equally affected by them.  A third
and more complex defect of the ideal graphene lattice is the
Stone-Wales deformation (SWd): Four hexagons are replaced by two
heptagon-pentagon pairs [see Fig.~\ref{fig:latt_defe}(c)]. As a
consequence, the \AB sublattice symmetry is broken, and \AB scattering
occurs.  To first order approximation, we adapt the tight-binding
parameters of the graphene ribbon to model the coupling parameters at
the SWd using geometry-dependent coupling parameters.


\begin{figure}[h]
\hbox{}\hfill\epsfig{file=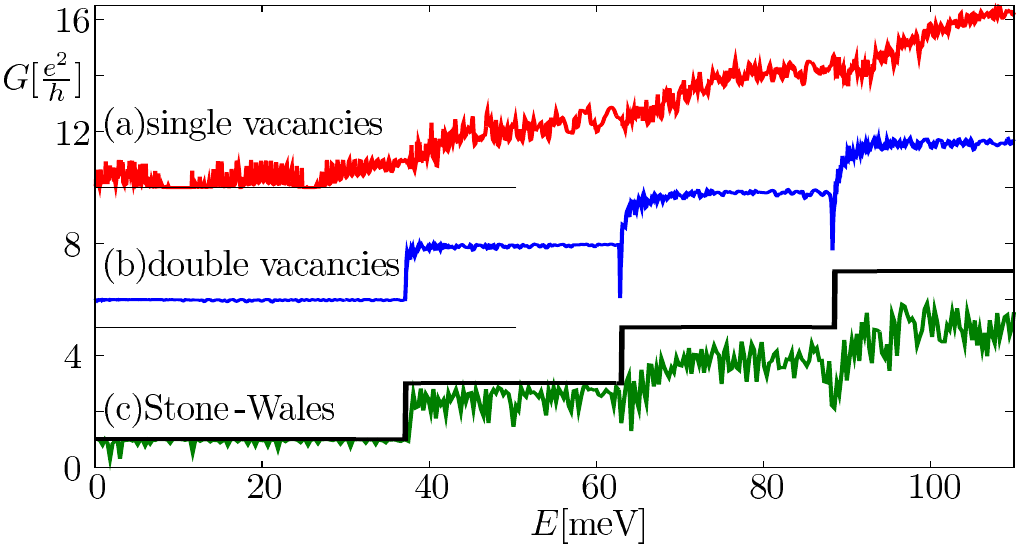,width=7cm}\hfill\hbox{}
\caption{(Color online){ Conductance of a 60 nm wide graphene
    nanoribbon with a length of 1 $\mu$m and a defect density
    $n_i=10^{-5}$ defects/atom of (a) single-point vacancies, (b)
    double vacancies, or (c) Stone-Wales deformations (curves
    vertically offset by 5$e^2/h$ for clarity). The staircase function
    for ideal size quantization plateaus is shown in (c) as thin black
    line.}  }
\label{fig_super}
\end{figure}

We perform quantum transport simulations for a zigzag graphene nanoribbon of
width $W\!= 60$ nm and length $L = 1 \mu$m in the presence of
disorder. ``Bulk'' disorder is introduced by randomly distributed
electronic lattice defects. Even relative defect concentrations as low
as $n_i = 10^{-5}$ defects/atom [Fig.~\ref{fig_super}] give rise to
pronounced deviations from the ideal staircase-shaped conductance $G$
with plateaus due to transverse size quantization. While both single-
and double vacancies lead to a reduction of transmission, the two
types of vacancies give rise to very different modifications.  While
pseudospin conserving double vacancies approximately preserve the
feature of quantization plateaus, with reduced height and pronounced
dips near the steps where additional modes open, the quantization
plateaus are completely washed out for single vacancies. This is all
the more remarkable as the total number of point defects for single
vacancies is only half the number for the double vacancies. This
clearly points to the breaking of the \AB sublattice symmetry as
origin of the loss of size quantization. The drastic difference
between pseudospin conserving double vacancies and non-conserving
single vacancies persists over a wide range of defect concentrations
and is robust against an average over many disorder configurations
(Fig.~\ref{fig:t_ni_e}).

\begin{figure}
\hbox{}\hfill\epsfig{file=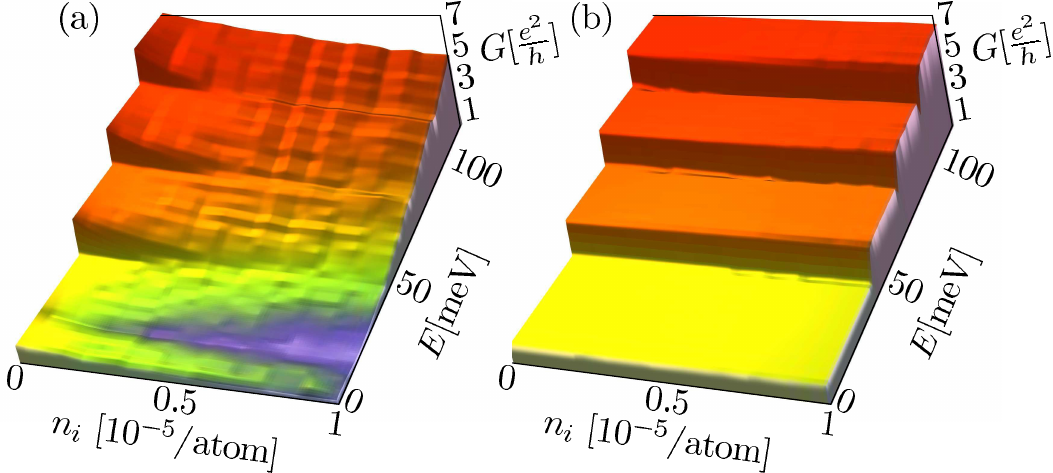,width=8.5cm}
\hfill\hbox{}
\caption{(color online) Conductance through a graphene zigzag
  nanoribbon of length 1 $\mu$m and width $W=60nm$ as a function of
  energy and disorder concentration $n_i$ (in units of $10^{-5}$
  defects/atom) averaged over 100 disorder realizations for (a)
  single vacancies and (b) double vacancies.}
\label{fig:t_ni_e}
\end{figure}

The connection between pseudospin conservation and transverse
quantization steps can be inferred from the relation between the
direction of the wave vector [Eqs.~(\ref{Eq:Thetak},\ref{pseudospin})]
and the helicity operator acting on the SU(2) representation-space
spanned up by the A and B sublattices. For the free Dirac equation
[Eq.~(\ref{Eq:H_Dirac})] the ratio between $k_x$ and $k_y$ is
determined by the pseudospin [see Eq.~(\ref{pseudospin})]. In turn,
transverse quantization for a finite-width ribbon relates the step
quantum number $n$ with $k_y$. Through the introduction of defects
that break pseudospin conservation, or interactions with an underlying
substrate, which invariably introduces a spatially varying electronic
environment $V(\x)$ [see Eq.~(\ref{Eq:H_Dirac})] breaking \AB
symmetry, the transverse quantum number $n$ becomes ill defined,
resulting in the strong suppression of transverse quantization
steps. Conversely, the absence of pronounced size-quantization
plateaus in experiment \cite{Kim2010,Molitor,lin08,LewenkopfReview}
hints at broken \AB symmetry in experimental structures. This
mechanism can be illustrated and verified with the help of our
quasi-classical trajectory simulations. Using either an angular
scattering probability at point defects that are pseudospin preserving
[Eq.~(\ref{proptozero})] or non-conserving [Eq.~(\ref{swavescatt})],
we can reproduce all qualitative features of quantum conductance in
remarkable detail (Fig.~\ref{fig:Semic}): for $s$-wave scattering,
quantization steps are strongly suppressed, giving an (approximately)
linear slope of transmission with energy. For pseudospin-conserving
scattering, quantization steps are pronounced, and the characteristic
dips in the transmission near the thresholds observed in the full
quantum mechanical calculation are reproduced. These dips resulting
from scattering into evanescent modes disappear in the case of
$s$-wave scattering, as the backscattered trajectory ``forgets'' its
new direction after the next scattering event. We have verified that
these striking differences between single- and double vacancy
scattering are not qualitatively changed by higher-order effects
(e.g.~triangular warping). The relation between broken \AB sublattice
symmetry and the destruction of size-quantization plateaus implies
that for disorder caused by different and more complex defects which
also result in pseudospin non-conserving scattering, signatures of
transverse quantization should also disappear. This can, indeed, be
verified for Stone-Wales defects [Fig.~\ref{fig:latt_defe}(c)]. The
size quantization plateaus are washed out by as few as 1 SWd in $10^5$
atoms [see Fig.~\ref{fig_super}(c)].
\begin{figure}[t]
\hbox{}\hfill\epsfig{file=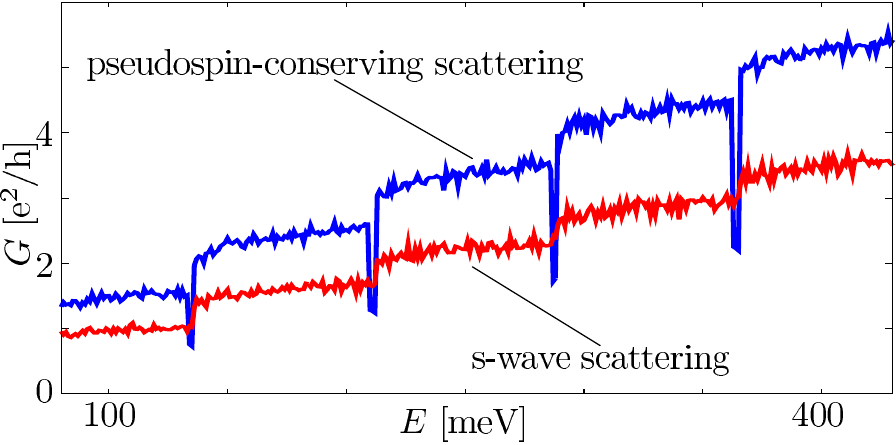, width = 8cm}\hfill\hbox{}
\caption{(Color online) 
Quasi-classical Monte-Carlo simulation of transport through a disordered nanowire,
using a Dirac-like linear dispersion relation, and either 
graphene pseudospin-conserving scattering (upper blue line) or
$s$-wave scattering (lower red line) at randomly distributed local defects.}
\label{fig:Semic}
\end{figure}

The role of pseudospin conservation in the persistence of transverse
quantization can also be extracted from the scattering wave functions
which we obtain from the full quantum calculation. In order to
identify structures in the scattering wave function as a function of
ribbon length, we average over the $y$ component of the scattering
wave function of a single incoming mode $n$
\begin{equation}
 \left< |\psi_n(x)|^2\right> = \frac 1W\int_0^W |\psi_n(x,y)|^2 \mathrm
 dy\label{norm:proj}.
\end{equation}
For disorder created by single-vacancy defects
[Fig.~\ref{fig:latt_scatt}(a)], the transmitted wave function in the
exit lead does not feature pronounced oscillations since it represents
a superposition of many transverse modes. By contrast, the scattering
states for a disordered nanoribbon with pseudospin conserving double
vacancies feature in this region well discernible oscillating patterns
on two length scales [see Fig.~\ref{fig:latt_scatt}(b)]: (i)
the short beating period of $\lambda = 0.7$ nm, corresponding to the
distance (in $k$ space) between the $\Gamma$ and $K$ point
(see Ref.~[\onlinecite{LibPRB08}]) and (ii) a much slower variation with a
length scale $\Lambda \approx 12$ nm [see
  Fig.~\ref{fig:latt_scatt}(b)] corresponding to the wavelength
associated with the linear dispersion relation $E=v_{\mathrm F} \hbar
k$ (i.e., the distance from the $K$ point to a given point on the Dirac
cone). These two length scales differ by almost two orders of
magnitude. The persistence of these two well-defined oscillatory
components supports the notion that the presence of disorder does not
destroy size quantization provided disorder scattering is pseudospin
conserving. 

To summarize, we have presented full quantum transport simulations
through disordered graphene nanoribbons of realistic size, and find
that even very low concentrations of defects that destroy the \AB
sublattice symmetry lead to destruction of transverse size
quantization plateaus. By contrast, randomly distributed defects that
preserve the \AB sublattice symmetry leave the quantization plateaus
intact while modifying the transmission function. Our present results
suggest that the difficulty in observing pronounced quantization
plateaus in the conductance can be motivated by to the presence of
pseudospin non-conserving defects. This finding suggests that the
variation in local electronic structure of graphene due to
interactions with an underlying substrate or with local chemisorbates
binding to individual carbon atoms could be sufficient to destroy size
quantization plateaus in the experiment
\cite{Molitor,Kim2010,lin08,LewenkopfReview}. Indeed, in agreement
with these results, very recent studies \cite{vanWees} demonstate that
plateaus of quantized conductance can be observed if graphene quantum
point contacts are both suspended from the substrate as well as
thoroughly annealed to reduce the number of adsorbates on the graphene
lattice.

\begin{figure}[h]
\hbox{}\hfill\epsfig{file=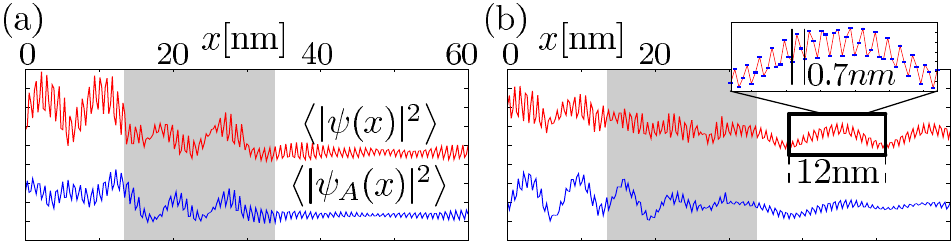,width=8cm}
\hfill\hbox{}
\caption{(Color online){ Averaged wave function density
    $\left<|\psi_1(x)|^2\right>$ [see Eq.~(\ref{norm:proj})] and
    sublattice density $\left<|\psi_{1,A}(x)|^2\right>$ of sub-lattice $A$
    for the scattering through a 10nm wide disordered graphene
    nanoribbon (disordered area shaded) featuring ten (a)
    single vacancy [(b) double vacancy] defects. }}
\label{fig:latt_scatt}
\end{figure}

We thank K.~Ensslin, J.~G\"uttinger, and C.~Stampfer for valuable
discussions.  Support by the Austrian Science Foundation (Grant
No.~FWF-P17359) and ViCoM SFB-041 is gratefully
acknowledged. Numerical calculations were performed on the Vienna
scientific cluster (VSC).  


\end{document}